\documentclass[usenatbib]{mn2e}

\usepackage{graphicx}
\usepackage{amsmath}
\usepackage{amssymb}
\usepackage{hyperref}
\usepackage{tablefootnote}

\usepackage{astrojournals}
\usepackage{doi}

\newcommand{\LCDM}{$\Lambda$CDM}

\newcommand{\rvir}{r_{\rm vir}}
\newcommand{\vmax}{V_{\rm max}}
\newcommand{\vin}{V_{\rm infall}}

\newcommand{\Lsun}{L_\odot}
\newcommand{\Msun}{M_\odot}

\newcommand{\slos}{\sigma_{\rm LOS}}
\newcommand{\kps}{km s$^{-1}$}

\newcommand{\nodata}{...}
\newcommand{\figrefname}{Fig.}
\newcommand{\figrefnameext}{fig.}

\usepackage[T1]{fontenc}
\usepackage{aecompl}

\title[M31 Satellite Masses in \LCDM]{M31 Satellite Masses Compared to \LCDM{} Subhaloes} 

\author[E. Tollerud et al.]{Erik J. Tollerud$^1$\thanks{Hubble Fellow}\thanks{$\!$erik.tollerud@yale.edu},
  Michael Boylan-Kolchin$^2$,
  James S. Bullock$^3$\\
  $^1$Astronomy Department, Yale University, New Haven, CT 06510, USA\\
  $^2$Department of Astronomy and Joint Space-Science Institute, University of Maryland, College Park, MD 20742\\
  $^3$Center for Cosmology, Department of Physics and Astronomy, The University of California, Irvine, Irvine, CA, 92697, USA}

\begin{document}

\label{firstpage}

\maketitle

\begin{abstract}
We compare the kinematics of M31's satellite galaxies to the mass profiles of the subhaloes they are expected to inhabit in \LCDM{}.  
We consider the most massive subhaloes of an approximately M31-sized halo, following the assumption of a monotonic galaxy luminosity-to-subhalo mass mapping.  
While this abundance matching relation is consistent with the kinematic data for galaxies down to the luminosity of the bright satellites of the Milky Way and M31, it is \emph{not} consistent with kinematic data for fainter dwarf galaxies (those with $L \lesssim 10^8 \Lsun$).
Comparing the kinematics of M31's dwarf Spheroidal (dSph) satellites to the subhaloes reveals that M31's dSph satellites are too low density to be consistent with the subhaloes' mass profiles.
A similar discrepancy has been reported between Milky Way dSphs and their predicted subhaloes, the ``too big to fail'' problem (TBTF).  
By contrast, total mass profiles of the dwarf Elliptical (and similarly bright) satellites are consistent with the subhaloes.  
However, they suffer from large systematic uncertainties in their dark matter content because of substantial (and potentially dominant) contributions from baryons within their half-light radii.

\end{abstract}


\section{Introduction}
\label{sec:intro}

Dwarf satellite galaxies of the Local Group (LG) are unique cosmological probes. They are close enough that very faint objects can be detected, and  their individual stars can be spectroscopically observed.  This enables measurements that are impossible with integrated-light observations \citep[e.g.,][]{sandg, stri08commonmass, walker09fnx}.

A problem revealed by these observations was recently pointed out by \citet{bkbk11,bkbk12}, dubbed the ``too big to fail'' problem.
They demonstrated that the bright satellites of the Milky Way (MW) have internal kinematics inconsistent with the  expectations of a \LCDM{} simulation of a MW-sized halo.  
Specifically, they showed that the most massive subhaloes in \LCDM{} simulations have central masses systematically larger than those measured in the ten brightest dSph satellites of the Milky Way. 
The population of satellites compatible with observed kinematics of the dSphs contains intermediate-mass subhaloes, leaving the question of why the most massive subhaloes appear to be dark.
This implies either that the massive subhaloes of the MW are inexplicably without luminous galaxies, or the central densities of their dark matter haloes are different from the expectations of a dark matter-only simulation in \LCDM{}. 

A number of explanations for this problem have been suggested, from forming cores in the galaxies' mass profiles to the existence of dark matter particles with unusual properties 
\citep{dicintio11, lovell12, zolotov12, vogelsberger12, vinas12, maccio13, rocha13, peter13, libeskind13, brooks13, dicintio13}.  
The simplest possibile explanation, however, is that the MW is an outlier relative to other similar galaxies. 
That is, if the MW satellite population were different from that of a typical galaxy with the same halo mass, it would cast doubt on the common practice of comparing the MW to typical halos in simulations.

The other bright spiral of the LG, M31, provides a second testing ground for this effect.  
It has a large population of known satellites, thanks in large parts to the efforts of the Pan-Andromeda Archaeological Survey \citep[PAndAS,][]{pandas09nat}.
While a few of its brightest satellites are classified as dwarf Ellipticals (dEs), most are dSphs, implying that the satellite system of M31 may be comparable to that of the MW. 
This (among other goals) has motivated spectroscopic surveys of the M31 satellites, including the dwarf component of the Spectroscopic and Photometric Landscape of the Andromeda Stellar Halo (SPLASH) survey \citep{kalirai10, howley12m32, paper1}, as well as a parallel survey by \citealt{collins13} (and a corresponding kinematics analysis in \citealt{collins13_2}, discussed more in \S \ref{sec:comp}). 
These and other kinematic investigations of M31 satellites provide a wealth of data for examining the internal dynamics of M31 satellites.

In this paper, we make use of these new data sets to compare M31's satellites to subhaloes from a \LCDM{} N-body simulation intended to approximate the expected halo of M31.  
In \S \ref{sec:sats}, we provide an overview of the M31 satellites and the observational data we use here.  
In \S \ref{sec:subs}, we describe the comparison simulation datasets and how we map them on to the observations. In \S \ref{sec:comp}, we compare 
the M31 satellites to the simulated subhaloes. Finally, in \S \ref{sec:conc}, we provide concluding thoughts. 
Where relevant, we assume a distance to M31 of 783 kpc \citep[e.g.,][]{paturel02, Mcconnachie05, perina09}.

\section{M31 Satellites}

\label{sec:sats}
\begin{figure}
\includegraphics[width=0.5\textwidth]{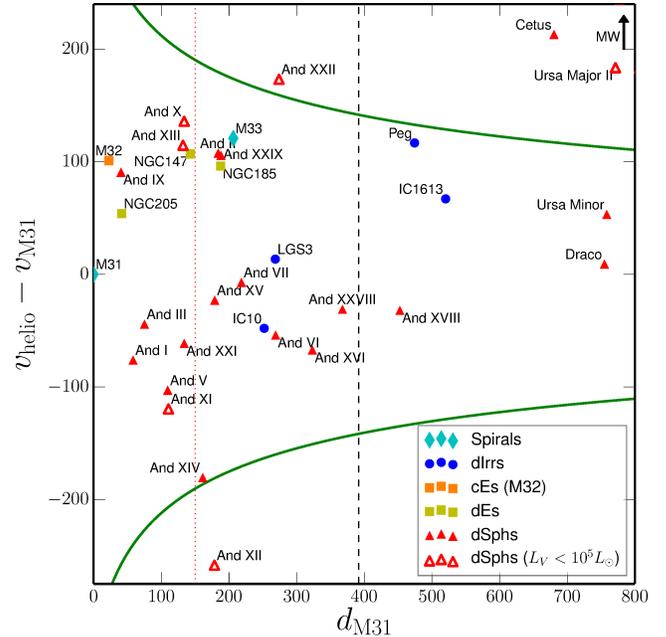}
 \caption{Local Group galaxies near M31 in real space and velocity space.  
 The velocity scale is line-of-sight heliocentric velocity relative to M31, while the distance axis is 3D distance from the centre of M31.  
 Spirals (M31 and M33) are cyan diamonds, dwarf irregulars are blue circles, dwarf ellipticals are yellow squares, the compact elliptical is an orange downward-pointing triangle, dSphs with $L_V> 10^5 \Lsun$ are filled red triangles, while those with $L_V < 10^5 \Lsun$ are open red triangles. 
  The black dashed vertical line indicates the distance halfway between the MW and M31, while the red dotted line indicates the distances out to which M31 may be complete due to the PAndAS survey.  
  The green curve indicates the one-dimensional escape velocity from an M31-like NFW halo (see text).  
  Data are from \citet{mcc12lgcat} and \citet{toll13}.
  This demonstrates that essentially all of the satellites near M31 are likely bound.}
 \label{fig:lgrv}
 \end{figure}

An important initial question is which galaxies near M31 should be treated as ``satellites.''  Because a goal in this paper is to compare the satellites of M31 to
 subhaloes of representative \LCDM{} dark matter haloes, it is reasonable to choose a definition that is informed by simulations of such haloes.  
 With this in mind, in \figrefname{}  \ref{fig:lgrv} we show line-of-sight velocity as a function of position for galaxies near M31.  The green curves in this figure are the escape velocity  divided by $\sqrt{3}$ for a 
 NFW \citep{NFW} halo  with parameters matching those suggested for M31 in \citet[][model $C_1$]{klypin02}. 
 These represent the \emph{average} line-of-sight  velocity of a population of subhaloes that are barely bound to M31 and have isotropic orbital velocities.  

 The fact that all but two of the galaxies in \figrefname{} \ref{fig:lgrv}  lie within this line indicates that they are likely \emph{all} bound to M31.  
 The two beyond the green line could still easily be bound, as they might simply have velocity vectors aligned close to the line-of-sight.  
 Thus, as a population, the satellites are consistent with being all bound.
 The results described in \citet{BK13} indicate that $\gtrsim 99.9 \%$ of subhaloes of a $\sim 10^{12} \Msun$ halo are bound to the halo.
 This implies that all of the galaxies with distances from M31 $\lesssim \rvir$ can be treated as satellites for comparing to \LCDM{} simulations.
 
 While this gives confidence that the galaxies on this plot are likely to be satellites (and thus inhabit subhaloes bound to their host), there is another 
 important constraint, indicated by the vertical black dashed line. This line indicates the distance half way from M31 to the MW.  Galaxies beyond this distance are difficult to 
 unambiguously assign to M31, because those that lie nearer to the MW may in fact be more influenced by the MW than M31.  This is indicated by the presence of 
 Ursa Minor and Draco in \figrefname{} \ref{fig:lgrv}, as their line-of-sight velocities are consistent with M31, but they are much closer to the MW than M31 and are universally regarded as 
 satellites of the MW.  Additionally, galaxies beyond this distance ``behind'' M31 (relative to the MW) become less likely to have interacted with M31, as it is not clear if they 
 are on wide orbits or first infall into the M31 system.  Because of this, in what follows, we focus on only those galaxies closer to M31 than 390 kpc as likely 
 M31 satellites.

 \figrefname{} \ref{fig:lgrv} also indicates the morphological type of galaxies near M31 by the type and colour of symbol.  It is immediately clear that the M31 satellite system includes 
 a variety of different galaxies types, in contrast to the MW (which has only dSphs and two dIrrs - the Magellanic Clouds). While this provides a rich environment for understanding
 a variety of processes in galaxy formation, it also complicates analysis of the M31 system as a whole, because the quite different morphologies of these galaxies
 necessitate different methods of measuring masses.  Below, we describe our methodologies for estimating masses for each of these categories of galaxies.  
 We also provide Table \ref{tab:sats} as a summary of the properties relevant for the satellites we consider in the following sections. 

\begin{table*}
\begin{minipage}{150mm}
\caption{Key Properties of M31 Satellites with $L_V > 10^5 \Lsun$.}
\label{tab:sats}
\begin{tabular}{@{}ccccccccc}
\hline

Name & 
Type & 
$\log(L_V / L_\odot)^{\rm a}$ &
$r_{1/2}^{\rm b}$ &
$V_c(r_{1/2})^{\rm c}$ &
$(M/L)_{\rm 1/2}^{\rm d}$ &
$\vmax^{\rm e}$ &
$\vin^{\rm f}$ &
Source$^{\rm g}$ \\
&&&[kpc]&[\kps]&$[\Msun/\Lsun]$&[\kps]&[\kps]& \\
\hline

And I & dSph & $6.7 \pm 0.4$ & $ 839 \pm 45$ & $18 \pm 4$ & $27 \pm 11$ & $14^{+3}_{-2}$ & $18^{+6}_{-4}$ & 1, 2 \\
And III & dSph & $6.0 \pm 0.1$ & $529.9 \pm 0.2$ & $16 \pm 2$ & $62 \pm 18$ & $14^{+3}_{-2}$ & $18^{+5}_{-4}$ &  1, 2 \\
And V & dSph & $5.8 \pm 0.1$ & $442 \pm 22$ & $18 \pm 3$ & $116 \pm 32$ & $17^{+4}_{-3}$ & $22^{+7}_{-5}$ &  1, 2 \\
And VII & dSph & $7.3 \pm 0.1$ & $972 \pm 44$ & $23 \pm 2$ & $13 \pm 3$ & $21^{+3}_{-2}$ & $26^{+8}_{-6}$ &  1, 2 \\
And IX & dSph & $5.2 \pm 0.2$ & $726 \pm 29$ & $19 \pm 4$ & $809 \pm 368$ & $15^{+3}_{-3}$ & $18^{+6}_{-5}$ &  1, 2 \\
And XIV & dSph & $5.32 \pm 0.03$ & $534 \pm 36$ & $9 \pm 2$ & $103 \pm 44$ & $11^{+1}_{-1}$ & $14^{+4}_{-3}$ &  1, 2 \\
And XV & dSph & $5.9 \pm 0.2$ & $355 \pm 39$ & $7 \pm 3$ & $11 \pm 9$ & $11^{+2}_{-1}$ & $14^{+4}_{-3}$ &  1, 2 \\
And XVI & dSph & $5.6 \pm 0.2$ & $179 \pm 18$ & $7 \pm 6$ & $9 \pm 7$ & $12^{+3}_{-2}$ & $15^{+5}_{-4}$ &  1, 2 \\
And XVIII & dSph & $5.80 \pm 0.03$ & $417 \pm 30$ & $17 \pm 5$ & $87 \pm 47$ & $13^{+3}_{-3}$ & $17^{+6}_{-4}$ &  1, 2 \\
And XXI & dSph & $5.66 \pm 0.03$ & $1023 \pm 70$ & $12 \pm 10$ & $159 \pm 132$ & $12^{+2}_{-2}$ & $15^{+5}_{-4}$ &  1, 2 \\
And XXVIII & dSph & $5.6 \pm 0.3$ & $282 \pm 54$ & $8 \pm 3$ & $22 \pm 16$ & $12^{+2}_{-2}$ & $14^{+4}_{-3}$ &  1, 2 \\
And XXIX & dSph & $5.6 \pm 0.2$ & $482 \pm 57$ & $10 \pm 2$ & $62 \pm 29$ & $11^{+1}_{-1}$ & $14^{+4}_{-3}$ &  1, 2 \\
NGC 147 & dE & $8.1 \pm 0.1$ & $364 \pm 24$ & $53 \pm 18$ & $4.2 \pm 0.6$ &\nodata & \nodata & 3\\
NGC 185 & dE & $8.2 \pm 0.1$ & $295 \pm 23$ & $52 \pm 19$ & $4.6 \pm 0.6$ &\nodata & \nodata & 3\\
NGC 205 & dE & $8.6 \pm 0.1$ & $520 \pm 29$ & $41 \pm 14$ & $1.0 \pm 0.7$ & \nodata & \nodata & 4\\
M32 & cE & $8.51 \pm 0.04$ & $110 \pm 16$ & $79 \pm 9$ & $1.0 \pm 0.2$ &\nodata & \nodata & 5, 6, 7\\
IC 10 & dIrr & $8.23 \pm 0.04$ & $612 \pm 39$ & $35 \pm 5$ & $2.0 \pm 0.6$ & \nodata & \nodata & 2, 8\\
LGS 3 & dIrr & $5.97 \pm 0.04$ & $626 \pm 63$ & $9 \pm 7$ & $23 \pm 19$ & $12^{+2}_{-2}$ & $15^{+5}_{-4}$ & 2, 9\\
M33 & SAcd & $9.45 \pm 0.04$ & $2344 \pm 297$ & $50 \pm 5$ & $1.0 \pm 0.2$ & $130^{+10}_{-10}$ & $130^{+10}_{-10}$ & 10, 2 \\

\hline
\end{tabular}

\raggedright
 $^a${Log of total V-band luminosity} \\
 $^b${Three dimensional (deprojected) half-light radius} \\
 $^c${Circular velocity at $r_{1/2}$: $V_c^2(r_{1/2}) = G M(<r_{1/2}) / r_{1/2}$} \\
 $^d${Mass-to-light ratio within $r_{1/2}$} \\
 $^e${Maximum circular velocity of galaxy's dark matter halo at $z=0$ (see \S \ref{sec:subs})} \\
 $^f${Maximum circular velocity of galaxy's dark matter halo at infall (see \S \ref{sec:subs})} \\
 $^g${References: 1: \citet{paper1}, 2: \citet{mcc12lgcat}, 3: \citet{geha10ngcs}, 4: \citet{geha06ngc205} , 5: \citet{howley12m32}, 6: \citet{choi02}, 7: \citet{galexcat07},  8: \citet{wilcots98},  9: \citet{cook99}, 10: \citet{simon06m33} } \\

\end{minipage}
\end{table*}

\subsection{Dwarf Spheroidals}
The non-starforming satellites of M31 are traditionally divided into dwarf Spheroidals (dSphs) and dwarf Ellipticals (dEs), 
approximately corresponding to a luminosity boundary of $L \sim 10^8 \; \Lsun$.  
Here, we follow this naming convention, because it conveniently separates the galaxies that are baryon-dominated from those that are not (\S \ref{sec:comp}). 
We do not use this terminology to imply anything specific about their formation or evolutionary histories, though.  For a more complete discussion of the elliptical-spheroidal dichotomy in such a context, see \citet{kormendy12}.  

The largest fraction of M31's known satellites are classified as dSphs, so they comprise the majority of our sample.  
These faint galaxies show little to no rotational support\footnote{with the exception of And II -- see \citealt{ho12}}, and hence their dynamical masses can be determined from their internal velocity dispersions \citep[e.g.][]{kalirai10, paper1, collins13}.  
Furthermore, as is clear from e.g. \citet{paper1} \figrefnameext{} 24c, they have mass-to-light ratios within their half-light radii that exceed  expectations for even the oldest stellar systems.  This implies that the dominant mass component is the dark matter, so the dynamical mass is an excellent proxy for the dark matter mass.  

For the dSphs, we therefore adopt the \citet{wolf10} mass estimator:
\begin{equation}
\label{eq:mhalf}
M_{1/2}^{\rm DM} \approx M_{1/2}(r_{1/2})=3 \frac{\slos^2 r_{1/2}}{G},
\end{equation}
where $\slos$ is the luminosity-weighted line-of-sight velocity dispersion, $r_{1/2}$ is the three-dimensional (deprojected) half light radius, and $M_{1/2}$ is the mass within that radius.  This mass estimator is valid with the given normalization only when interpreted as the mass within this particular radius\footnote{The mass estimator also assumes equilibrium and spherical symmetry, and that the velocity dispersion profile is flat with radius.}.  If interpreted in this way, it is insensitive to the galaxy's stellar velocity dispersion anisotropy.  
As discussed in detail in \citet{wolf10} Appendix B, $r_{1/2}$ can be estimated at the $2$ percent level as a simple scaling of $R_{\rm eff}$ for light profiles like those of the MW dSphs \emph{or} most other light profiles that are at all plausible for galaxies. 

We apply these estimators to the compilation of M31 kinematics and luminous properties from \citet{paper1}, and include these as the dSph satellites in Table \ref{tab:sats}.  Translating the above mass estimator to circular velocities ($V_c^2 = G M / r$) at the 3D half-light radii yield the results described in the following sections.  We note that a comparable data set exists in \citet{collins13}, and while we do not include those data here, they are generally  consistent where the data sets overlap, and show similar scalings. 

A final consideration particular to the dSphs is that of exclusion based on luminosity. While the other categories of satellites contain only relatively bright 
satellites, dSphs extend to the detection limits of the PAndAS survey \citep{brasseur11}. The presence of  such satellites around the MW that are fainter suggests that M31's dSph population continues below the current luminosity threshold for luminosity.  
Furthermore, faint dSphs around M31 and the MW  likely suffer from biases due to surface brightness \citep[e.g.][]{b10stealth, brasseur11, martin13}.  
Therefore, we adopt the criterion $L_V>10^5 \Lsun$ to match the luminosity at which both MW and M31 dSph samples are likely unbiased in luminosity.
This also corresponds roughly to the luminosity at which larger (lower surface brightness) galaxies are absent in the MW, suggesting that most, if not all, of the galaxies above this limit are detected \citep{brasseur11}.
The dSphs accepted by this criterion are shown in \figrefname{} \ref{fig:lgrv} as filled red triangles. This sample thus represents the \emph{brightest} dSphs in the M31 system (within 150 kpc - see the discussion in \S \ref{sec:comp}). 

\subsection{Dwarf and Compact Ellipticals}

As described in the previous section, the dE satellites of M31 are the non-starforming satellites above a conventional luminosity boundary  of 
$L \sim 10^8 \; \Lsun$.   While it is plausible that they represent an evolutionary continuum with dSphs, this dichotomy does have some value in that the 
dEs have much higher central surfaces brightnesses.  As a result, dSphs can generally be considered dark matter dominated, while dEs cannot.  
Furthermore, while the dSphs  are primarily dispersion supported, the  dEs can have substantial rotation \citep[e.g.][]{geha06ngc205,geha10ngcs,howley12m32, paper1}.
These factors combine to necessitate a different approach to mass estimation than that described above for the dSphs.

The two fainter M31 dEs NGC 147 and NGC 185 are quite similar in luminosity, morphology, and kinematics.  Our mass estimate for them is from \citet{geha10ngcs}, 
which made use of dynamical modelling to fit velocities of hundreds of resolved stars in each galaxy.  These models make simplifying assumptions to improve their
stability, one of the most relevant of which is that mass-follows-light (i.e., a constant mass-to-light ratio).  While this is not in detail accurate in a \LCDM{} context where
the dark matter haloes and baryonic components have different profiles, such modelling should provide a reasonable mass estimate when the baryons contribute 
a substantial portion of the mass budget.  With this approach, \citet{geha10ngcs} find mass-to-light ratios that are not consistent with a \emph{purely} stellar 
component , but the stars provide the dominant mass within the half-light radius.
This signifies the presence of dark matter, but also indicates that these galaxies are baryon-dominated in the central regions.
In the absence of a purely general two-component mass model for these galaxies, we adopt the dynamical mass estimate of \citet{geha10ngcs} as the total mass for these dEs.

While not strictly a dE, we adopt a similar mass estimate for the unusual M31 satellite M32.  It is similar in luminosity to the M31 dEs, 
but has a much smaller size, and hence higher surface brightness. This places M32 in the rare (unique in the LG) class of ``compact elliptical'' (cE).   
Resolved-star spectroscopy is impossible in its central regions due to the high level of crowding. 
Such measurements are possible in its outskirts, however, as shown in \citet{howley12m32}. By combining these observations with slit spectroscopy
of the central regions, it is possible to construct a dynamical model of M32.  While this model must account for the nuclear black hole \citep{vdm98, joseph01}, it produces dynamics consistent with purely stellar populations.  While M32 may well contain dark matter, its small size
and high luminosity combine to make measurement of any dark matter halo it may have nearly impossible.  Accordingly, we simply treat the baryonic mass found in
\citet{howley12m32} as an upper limit on its dark halo mass.

The last dE satellite of M31 is NGC 205.  While outwardly similar to the other dEs, is complicated by the presence evidence of ongoing tidal disruption 
\citep[e.g.][]{choi02, geha06ngc205}.  It also has a somewhat lower $v/\sigma$ than the other dEs, indicating more pressure support.  Because of this,
we adopt the mass estimator of Equation \ref{eq:mhalf} using the global dispersion for the resolved star velocities tabulated in  \citet{geha06ngc205} inside the radius 
at which the velocity curve turns around.  While the lack of equilibrium indicated by the tidal disruption violates the assumptions used to derive Equation \ref{eq:mhalf},
the use of only stars that do not show direct evidence of being affected by tides provides some hope that the estimator can provide at least a very rough bound on the
mass of NGC 205.  However, as is the case for the other dEs, the high luminosity implies that much of the mass this estimator implies can be provided by
the stars.  Thus, like for NGC 147 and 185, our mass estimate can only be treated as an upper bound on the \emph{dark} mass of these galaxies within their
half light radii.

\subsection{Dwarf Irregulars}

Moving beyond most of the quiescent satellites of M31, we find the M31 satellites LGS 3 and IC 10.  
These galaxies have recent star formation and detectable gas, placing them conventionally in the category of dwarf Irregular (dIrr) galaxies \citep{roberts62, thuan79, wilcots98, bouchard06}.  They do, however, inhabit quite different luminosity regimes, as LGS 3 is more comparable to the 
dSphs ($M_V \sim -10$, $L_V \sim 10^6 \Lsun$), while  IC 10 is more akin to the dEs ($M_V \sim -15$, $L_V \sim 10^8 \Lsun$).

Because of the presence of neutral hydrogen, resolved observations of the HI spectral line in IC 10 provide the possibility of a rotation curve that can provide a rough mass estimate.  
Such a mapping in \citet{wilcots98} revealed that, while a rotating disc is present, it is strongly perturbed by holes and shells driven by the intense current star formation. 
Thus, while the HI rotation curve provides a method of mass estimation for IC 10 that is distinct from the dSphs and dEs, the mass can only by estimated at a 
precision comparable to that of the other satellites.  Furthermore, as with the dEs, its luminosity is high enough that the stellar component's mass is comparable to, if
not great than, that of the dark matter.  Thus, as with the dEs, we treat the mass derived from the \citet{wilcots98} observations of IC 10's HI to be an upper
limit for the dark matter mass of IC 10.

By contrast, while LGS 3 does have associated HI clouds likely associated with it \citep{bouchard06, grcevich09}, the clouds are not yet resolved well enough to measure
a rotation curve and infer a mass.  Therefore, for LGS 3, we fall back on the radial velocity data set of \citep{cook99}, following the same approach as for the dSphs.  
We caution that the observations of \citet{cook99} include only four LGS 3 member stars, so the velocity dispersion is subject to very large uncertainty.  
However,
this is as yet the only available appropriate dataset for this galaxy, so we use this dispersion-based mass to provide a rough mass estimate for LGS 3. 
This estimate reveals that, even with the lowest mass accepted by the error bars, LGS 3 has a large mass-to-light ratio akin to the dSphs, and we therefore include it in our sample of  dark matter mass estimates.

\subsection{M33}

The most luminous galaxy in the M31 system aside from M31 itself (and the third spiral of the LG) is M33.  While an interesting object to study on its own merits, 
here we are primarily concerned with it as a satellite of M31.  \figrefname{} \ref{fig:lgrv} reveals that it is plausible to treat it as such in a \LCDM{} context, as it is clearly less massive 
than M31, with a relative velocity implying it is bound to 
M31.  
This is further supported by proper motions and dynamical analysis of the M31-M33 system \citep{brunthaler05, vdm12}, and the suggestion that it has tidally interacted with M31 \citep{mcc10m33}. 

M33 is particularly valuable for our analysis here in that it has a measurable HI rotation curve that extends well beyond its optical extent \citep[e.g.][]{corbelli03}.
This clearly reveals the presence of a dark matter halo, and allows fitting of the baryonic and dark halo components.  For consistency with 
the other satellites, we adopt only the dark halo component within the half-light radius as our mass estimate for M33, 
based on the combined HI \citep{corbelli03} and 2MASS \citep{jarrett03} analysis of \citet{simon06m33}.

\section{Subhalo Sample and Abundance Matching}
\label{sec:subs}

The previous section described a well-defined sample of M31 satellites with mass estimates (or at least upper limits) within a specified three-dimensional distance.  
We now turn to defining a sample of subhaloes in a \LCDM{} context suitable for comparison with these satellites. 
A natural starting point for this connection is the abundance matching technique \citep{valeost04, krav04, conroy06}.  
Abundance matching depends on the assumption that the (sub)halo mass to galaxy luminosity (or stellar mass) relation is monotonic.  
This assumption allows the straightforward procedure of mapping a galaxy's luminosity on to a dark matter halo mass by simply matching cumulative number densities of galaxies to haloes.  
This approach has been shown to be consistent with a variety of observables for the scales at which luminosity functions can be reliably measured \citep[e.g., ][]{tg11}.

  \begin{figure}
\includegraphics[width=0.53\textwidth]{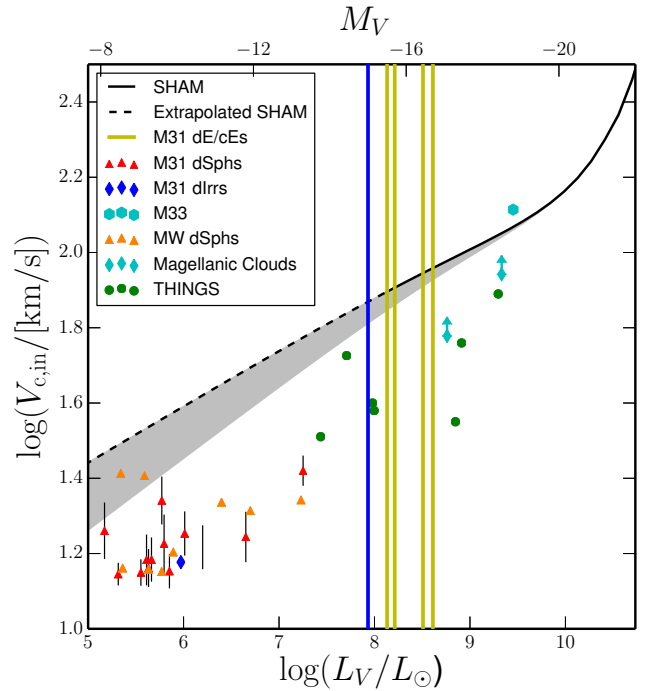}
 \caption{The luminosity-halo circular velocity relation of dwarf galaxies compared to (extrapolated) abundance matching.  
 The solid line is the abundance matching relation constructed from the \citet{baldry12} stellar mass function as described in the text. 
 The dotted line is constructed by extrapolating to lower luminosities using a simple power law.  
 The grey shaded region represents the possible change to this relation if a correction for surface brightness is included using the estimates from \citet{blanton05} (see text).
 The points are the circular velocity at infall for dwarf galaxies of the LG, estimated following the procedure described in the text.
 The triangles are dSphs, orange from the MW \citep{tollerud11a},  and red from M31, with error bars shown as a combination of observational errors and the intrinsic scatter in the concentration-mass relation.  
 Yellow vertical lines represent the M31 dEs and M32 - they are shown only as lines because there is too much uncertainty in their observed dark matter mass to constrain their halo mass.  The blue diamond and blue vertical line are M31 dIrrs,  the teal diamonds are the Magellanic Clouds, and M33 is the teal hexagon.
 The green circles are field dwarf galaxies from THINGS \citep{oh11things}.  
 It is clear from this that the dwarf galaxies do not match abundance matching nor its extrapolation, but \emph{do} have monotonically increasing halo mass with increasing luminosity.
   }
 \label{fig:vinvsl}
 \end{figure}

In \figrefname{} \ref{fig:vinvsl} we compare the halo mass-luminosity relation of this abundance matching approach to the satellites in our data set (\S \ref{sec:sats}).
We obtain the black line by performing abundance matching between the \citet{baldry12} stellar mass function and the (sub)halo mass function of the Millennium and Millennium II  simulations, assuming a uniform $M/L = 2$ \citep[see also][]{garrisonk13}.  
At $L_V  \lesssim 10^8 \Lsun$, the \citet{baldry12} stellar mass function becomes both incomplete and poorly-sampled due to detection limits, and hence beyond that point we depict the relation as a dotted black line and simply extrapolate the $\alpha = -1.47$ faint-end slope to arbitrarily low luminosities. 
The line does, however, seem to be roughly what is necessary to reproduce the luminosity function of LG dwarfs in \LCDM \citep{garrisonk13}.

Surface brightness incompleteness is not accounted for in the \citet{baldry12} stellar mass function.  
Substantial incompleteness on the faint end would mean the real universe has more faint galaxies than we assume here, and thus the correct abundance matching relation should have lower halo masses at fixed stellar mass.  
To estimate the possible magnitude of this effect, we also show (as the grey shaded region) the area that might be covered if the faint-end slope is as steep as $\alpha = -1.58$.
This faint-end slope corresponds to adding the amount of surface brightness correction \citet{blanton05} infers is necessary for the Sloan Digital Sky Survey.  
While this is not correct in detail for our relation (and does not reproduce the LG luminosity function, according to \citealt{garrisonk13}), we use it here simply as an estimate of the extent to which surface brightness incompleteness might affect our extrapolated abundance matching relation.

To compare with this abundance matching relation, \figrefname{} \ref{fig:vinvsl} also has an inferred circular velocity at infall ($\vin$) for our data set of M31 satellites.
We show the $\vin$ values of the M31 satellites as red triangles, yellow squares, and the teal hexagon.  
To determine these values, we follow the procedure outlined in \citet{bkbk12} to constrain $\vmax$ and $\vin$ for the dwarf satellite population of M31. Specifically, we compute circular velocity profiles of subhaloes across the 6 Aquarius simulations \citep{springel08aqrsubs} and compare these with the measured masses (at $R_{1/2}$) of the M31 dwarfs. 
For further details on this procedure, see \citet{bkbk12}. 
We note that this technique uses the particle data directly, and makes no assumption about the functional form of the density profile. 
It \emph{does} require assuming that low-mass satellite galaxies inhabit subhaloes with mass profiles like those found in dissipationless cosmological simulations, an assumption we aim to test with this data set.

For context, we also show the masses of MW satellites from \citet{tollerud11a} and \citet{bkbk12} as blue triangles and teal diamonds, as well as a sample of field dwarf galaxies from THINGS \citep{oh11things} as green circles. 
The Magellanic Cloud data points are lower limits, as they are based on present-day masses (and the Clouds may have had some of their mass stripped). 
We also show vertical lines at the luminosities of the M31 dEs and IC 10. As discussed in \S \ref{sec:sats} and \ref{sec:comp}, these satellites' kinematics are baryon-dominated, so their dark matter masses are too uncertain to reliably estimate $\vin$.

It  is immediately clear from \figrefname{} \ref{fig:vinvsl} that the dwarf galaxies are not consistent with naively extrapolating the abundance matching mass-luminosity relation.  
While the Magellanic Clouds and M33 may not be discrepant, the lower luminosity galaxies  have masses increasingly further below the extrapolated relation as luminosity 
decreases.  This holds even for the steepest extrapolated faint-end slope we consider here.  This suggests that the luminosity-halo mass relation (and thus galaxy formation) has an additional feature that sets in at luminosities below that of the Large Magellanic Cloud (LMC, $L \sim 10^9 \Lsun$). 
Further, this feature likely involves making haloes ``dark'', because if the relation of \figrefname{} \ref{fig:vinvsl} went through the points, the implied luminosity function is inconsistent with the observed luminosity function of the LG \citep{garrisonk13}.

\section{Comparing Satellites to Subhaloes}
\label{sec:comp}

\begin{figure}
\includegraphics[width=0.49\textwidth]{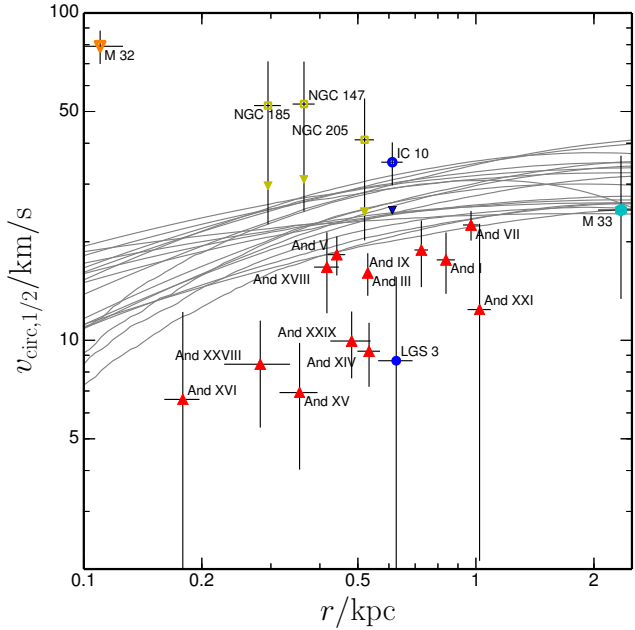}
 \caption{Circular velocities of M31 satellites and subhalo profiles.  
 The subhalo mass profiles are represented as the grey lines, and are selected as the $n$ most massive subhaloes where $n$ is the number of satellites shown. 
 The points show the mass within the half-light radius for the M31 satellites, with error bars.  
 Red triangles are dSphs, yellow squares are dEs, the orange downward-pointing triangle is the cE, blue circles are dIrrs, and the teal diamond is M33.  
 For the dEs, M32, and IC 10, the points are open symbols, as the measurements are upper limits, with no subtraction of the baryonic mass component.
 Also shown for the dEs and IC 10 are estimates of baryon-subtracted masses, although we emphasize these are quite uncertain.
 See text in \S \ref{sec:sats} for details of how masses are estimated, and \S \ref{sec:comp} for analysis of the patterns apparent in this figure.}
 \label{fig:tbtf}
 \end{figure}

We have defined a sample of observed M31 satellites with mass estimates at a particular radius (\S \ref{sec:sats}), and 
have also described a comparison sample of subhaloes from a plausible choice for an M31-like halo (\S \ref{sec:subs}). 
We now compare the kinematics of the observed satellites to the circular velocity profiles for the subhaloes in the simulations. 
\figrefname{} \ref{fig:tbtf} provides this comparison.  
The 19 satellites of M31 considered in this paper are shown as points with error bars  (symbols match those of \figrefname{} \ref{fig:lgrv}), and the grey lines show the circular velocity profiles of 19 subhaloes.  
Specifically, the curves correspond to the 19 most massive subhaloes of the Aquarius E halo (computed using the particle data directly).  
This number of subhaloes was chosen to match the number of observed satellites in our sample.

The Aquarius E halo has a virial mass of $1.4 \times 10^{12} \Msun$, which is in the middle of the range of recent estimates for M31 \citep{watkins10,paper1,vdm12,fardal13}. Choosing a higher (lower) mass host would result in more (fewer) subhaloes at a given value of $\vmax$. 
Further, the halo has an assembly history that is typical for haloes of its mass \citep{bk10nplh}. 
This means that we should expect our mapping of subhaloes to satellites to yield similar numbers of subhaloes and satellites.  For the 19 most massive haloes in the Aquarius E halo, $v_c \gtrsim 25$ \kps.
Abundance matching (\S \ref{sec:subs}) maps this subhalo $v_c$ threshold to a luminosity threshold close to what we use for the M31 satellites ($L_V> 10^5 \Lsun$).  
This means the counts are consistent with the M31 satellites, as a population, lying within the correspondingly most massive \LCDM{} subhaloes.

A few features are immediately apparent from \figrefname{} \ref{fig:tbtf}.  First, the dynamics of the dEs, M32, and IC10 imply densities too high for the subhaloes they are expected to inhabit. 
However, all of our measurements for these galaxies are \emph{upper} limits on the mass, as  the points assume all of the mass is dark. 
In practice, their baryonic components contain non-negligible mass, which would lower the dark matter contribution to the rotation curve, and that is the comparison of relevance to the curves on the plot.

To estimate the impact of the baryons' mass, we include down-pointing triangles for  dEs and IC 10 indicating where they would be if their stellar component is subtracted.  
For the dEs, this baryonic mass is estimated from the stellar populations inferred by \citet{geha06ngc205} and \citet{geha10ngcs}, and for IC10 this is done by simply assuming a fixed $M/L=2$. These points fall within the subhalo circular velocity profiles, implying that their kinematics are consistent with the subhaloes they are expected to inhabit. For M32, this is impossible to estimate, as M32 is consistent with the mass being provided by a purely stellar population.  \figrefname{} \ref{fig:tbtf} shows this is not surprising, as M32 is so compact that the dark matter mass of even the most massive subhalo would not be detectable over such a small volume.

However, it is important to recognize that this baryonic mass correction could be much larger than estimated above for stars with $M/L = 2$, as in all cases the baryonic component is at least half of the total mass (see Table \ref{tab:sats}). This means that systematic errors in the stellar mass estimates at the $50\%$ level imply \emph{any} dark matter mass below the open point in \figrefname{} \ref{fig:tbtf} is consistent with the data. Furthermore, baryon-dominated haloes may suffer adiabatic contraction or other interactions between the dark and baryonic components that distort the dark matter mass profile.  These effects are not accounted for in the dissipationless simulations, adding further uncertainty to the comparison between the subhalo profiles and kinematics of the bright satellites.

As a result, we conclude that while the observations are formally compatible with living in \LCDM{} haloes, the influence of the baryons is strong enough that we simply cannot meaure their dark matter content.  
That is, the likelihood that they are consistent with the halos is about the same as the likelihood that they are \emph{more} discrepant than the dSphs (see next paragraph).
In contrast to the ambiguous dEs, M33 \emph{is} compatible with the subhalo $v_c$ profiles, but it is unique among M31's satellites in being amenable to dark matter measurements from its rotation curve.
These same galaxies are discussed by \citet{veraciro13}, who find a consistency with \LCDM{} subhaloes for a sample composed primarily of these bright M31 satellites, but do not analyze the fainter dSphs. 

The other point clear from \figrefname{} \ref{fig:tbtf} is that the dSphs (and LGS 3) are systematically less dense than the most massive subhaloes. 
While the most massive satellites are marginally consistent with the least massive subhaloes, as a population they are clearly too low in density  to be consistent with the density of the subhalo abundance matching implies they should inhabit.  
This inconsistency with the naive expectations of \LCDM{} is remarkably similar to that of the bright satellites of the MW \citep{bkbk11,bkbk12}: the TBTF result discussed in \S \ref{sec:intro}.  
Taken together with the results on M33 and the dEs discussed above, it seems the TBTF effect becomes relevant somewhere between the mass of M33 and the mass of the brightest dSphs of M31.

 The interpretation of this result is complicated by the fact that the census of M31 satellites is likely incomplete beyond $150$ kpc due to the limited extent of the PAndAS survey.   PAndAS extends to $\sim 150$ kpc at the distance of M31, with additional coverage near M33 \citep{pandas09nat, brasseur11, yniguez13}. This includes around half of the virial volume, primarily in the central regions.  It is expected that subhaloes with larger $\vin$ are more likely to host bright satellites, and larger $\vin$ subhaloes tend to be closer to the host \citep{madau08, tollerud08, font11}.  So it is plausible that the majority of satellites are within the PAndAS footprint.  However, it is likely that at least some satellites remain to be discovered in the outer reaches of the M31 halo. This is further supported by the existence of faint dSphs such as And XXIX, XXXI, and XXXII, which are beyond the PAndAS coverage on the sky but are likely M31 satellites \citep{Bell11And29, toll13, martin13andsats}. 
 
\begin{figure}
\includegraphics[width=0.49\textwidth]{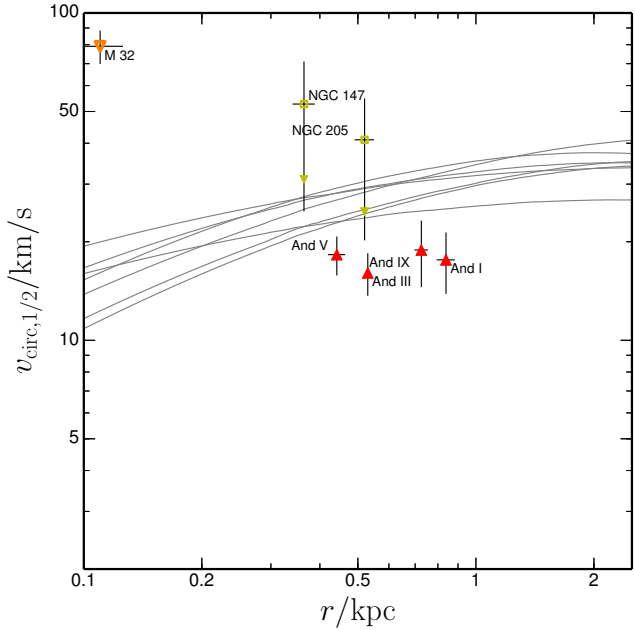}
 \caption{Same as \figrefname{} \ref{fig:tbtf}, but for satellites within 150 kpc of M31, the distance at which samples are guaranteed to be complete with respect to sky coverage. 
 The patterns here are consistent with those from \figrefname{} \ref{fig:lgrv}, although the small number of objects limit the utility of this subsample.}
 \label{fig:rm150}
 \end{figure}

 If the observed census of satellites is indeed strongly incomplete, we would need to compare some subsample of the subhaloes to the observed satellites.  
  Because we rank-order the subhaloes by mass, adding more subhaloes might lead to some  of the subhalo mass profiles overlapping with the observed points in \figrefname{} \ref{fig:tbtf}.  
  To test for this effect, in \figrefname{} \ref{fig:rm150}, we show a comparable plot to \figrefname{} \ref{fig:tbtf}, but only for M31 satellites and subhaloes that are within $150$ kpc.
 This distance is also convenient for comparison with the MW, as it is roughly the radius at which the M31 and MW satellite radial profiles seem to diverge \citep{yniguez13}.
 \figrefname{} \ref{fig:rm150} shows that the same trends hold as in \figrefname{} \ref{fig:tbtf}: the dSphs are lower density than the subhaloes they should inhabit, while the dEs are \emph{consistent} with the subhaloes (although they are only upper limits). 
 While the number of satellites is much smaller, this provides some confidence that the results of \figrefname{} \ref{fig:tbtf} are not strongly influenced by incompleteness in the surveys.  

The absence of a trend with distance from M31 is also relevant on its own.
Such a trend might be expected if tidal forces lower the central densities of satellites \citep[e.g.,][]{penn09}.  
While this effect is partially accounted for by the tidal effects built into the simulations we use here, collisionless simulations do not form stellar discs, and these might enhance the tidal effects \citep{penn10, brooks13}.
However, the fact that there is no trend between the TBTF effect and distance from M31 suggest that these effect are sub-dominant to the underlying cause of the TBTF phenomenon.

 While this manuscript was being completed, a similar analysis of the kinematics of M31 satellites \citep{collins13_2} appeared on the arXiv. 
 This study focused only on the dSphs of M31 (and the MW) using a similar dataset but distinct dataset to that used here.  They reached similar conclusions about the present-day masses of the M31 dSph (and the kinematic data are consistent), but their interpretation of the result is somewhat different.

\subsection{Impact of M31 Satellites on Interpretations of TBTF}

A variety of interpretations have been offered to explain the lower than predicted densities of the MW dSphs.  
These solutions range from the interactions of baryons with the satellites' dark matter haloes \citep[e.g.,][]{dicintio11, zolotov12, brooks13, dicintio13} to changes in the nature of the dark matter \citep[e.g.,][]{lovell12, vogelsberger12, vinas12, maccio13, rocha13, peter13, libeskind13}.
Our findings here for the M31 satellites are additional data to test such mechanisms, as they include a larger data set of dSphs around a second host, and demonstrate that the observations are not as constraining for satellites in the dE luminosity regime.

Additionally, the M31 dataset provides a direct constraint on the explanations that depend on the MW holding outlier status, either in satellite populations or in mass \citep[e.g.,][]{purcell12, wang12}.
The simplest explanation for the MW results might be that the MW is simply a statistical fluke, with a relatively small number of massive subhaloes.  
In that case, the MW dSphs could be consistent with the data, as they would have on average lower subhalo masses than a ``typical'' halo of the same mass ($M_{\rm vir} \sim 1-2 \times 10^{12} \Msun$), and thereby be consistent with the observations.  For example, \citet{purcell12} suggest this may occur for $\sim 10\%$ of such haloes.  The addition of the M31 dSph data set renders this explanation less probable, as it now requires \emph{two} such satellite systems, a $1 \%$ event (if the M31 and MW satellite  populations are uncorrelated, as the results of \citealt{garrisonk13} imply).

A similarly straightforward interpretation is to explain the TBTF effect by lowering the mass of the MW.  
If the MW's halo is somewhat lower than typically assumed \citep[$M_{\rm vir} \sim 10^{12} \Msun$: e.g., ][]{klypin02}, the typical subhalo mass is lower.
While this does not alter the abundance matching results discussed in relation to \figrefname{} \ref{fig:vinvsl}, it would alleviate TBTF around the MW.
That is, if the MW's halo mass is low, the typical mass of a subhalo hosting a MW satellite is lower, and this can render the dynamics of MW satellites consistent with the subhaloes  \citep{wang12}.  

However, the M31 results strongly constrain this argument.  Recent results for the proper motion of M31 and M33 by \citet{vdm12} coupled with the timing argument constrain the mass of the LG to $3.17 \pm 0.57  \times 10^{12} \Msun$.  Because the mass of the LG is dominated by the MW and M31, at least one of them must be $\gtrsim 1.5 \times 10^{12} \Msun$.  Because they \emph{both} show low densities for their dSphs, at least one of them must be massive enough to be in tension with their expected subhalo populations.  The fact that the satellite population of M31 seems to be larger (at fixed luminosity) than the MW further suggests that M31 may be the more massive in the pair\footnote{This assumes the M31 and MW together are statistically typical in their accretion history, which is certainly not guaranteed for only two galaxies.} \citep{yniguez13}.

\section{Conclusions}
\label{sec:conc}

In this paper, we have compared the internal dynamics of M31's  satellites  (with $L_V > 10^5 \Lsun$)  to the subhaloes they are expected to inhabit in a  \LCDM{} universe, given a simple set of assumptions for mapping galaxies to haloes. 
This yields the following main results:

\begin{enumerate}

\item The bright satellites of both M31 and the MW are consistent with a monotonic luminosity-to-$\vmax$ relation, the primary assumption necessary for abundance matching.  However, they are \emph{not} consistent with extrapolating the abundance matching of $\gtrsim 0.1 L_* $ galaxies to lower luminosities.
Specifically, low-luminosity dwarfs are significantly less dense than an abundance matching relation matching the LG luminosity function would predict.  
These results are robust to assumptions about the masses of the M31 and MW halo.

\item The dEs and similarly bright satellites of M31 (which have no analogue in the MW) are nominally consistent with the subhaloes they would be expected to inhabit.  
However, the interpretation of this in the context of their dark matter haloes is hampered by the significant contribution to their mass budget by  their baryonic mass and the associated possible systematic uncertainties.

\item The dSph satellites of M31 have lower densities (within their half-light radii) than the densities  of the most massive subhaloes that are expected to host them in \LCDM{} collisionless dark matter simulations.
Their dynamics are thus consistent with the satellites of the MW, and exhibit the ``too big to fail'' problem. 
Thus, the simplest explanation for the problem -- that the MW is a statistical fluke  -- does not seem to be valid, as a statistical anomaly is much less likely to be found in two different haloes.

\end{enumerate}

These results provide crucial context for interpreting the small-scale puzzles presented by \LCDM{} or alternative cosmological models by moving beyond the MW and its satellites. 
They demonstrate that these puzzles persist in a very similar form for M31 and its satellite galaxies.
However, understanding if they also hold for a larger, statistical sample of galaxies is necessary for interpreting such results in a cosmological context.
Unfortunately, resolved star spectroscopic data like those used here are nearly impossible to obtain beyond the Local Group with current spectroscopic capabilities.
Fortunately, in the coming era of deep, wide surveys like the Large Synoptic Survey Telescope and extremely large telescopes for spectroscopic follow-up (e.g., the Thirty Meter Telescope and Giant Magellan Telescope), the potential exists to push the boundaries of near-field cosmology well beyond the Local Group.

\vskip 1.5\baselineskip

\noindent {\bf{Acknowledgements}} 

The authors thank Marla Geha, Ana Bonaca, Jorge Pe\~narrubia, Alis Deason, Andrey Kravtsov, Carlos Frenk, and the anonymous referee for valuable discussions.  
We also thank the Aquarius collaboration for providing access to their simulation data.

This research made use of Astropy, a community-developed core Python package for Astronomy \citep{astropy}. 

Support for this work was provided by NASA through Hubble Fellowship grant \#51316.01 awarded by the Space Telescope Science Institute, which is operated by the Association of Universities for Research in Astronomy, Inc., for NASA, under contract NAS 5-26555. This work was also supported by NSF Grant AST-1009973. 

\bibliography{andtbtf}{}
\bibliographystyle{hapj} 


\end{document}